\documentclass[11pt, a4paper]{article}
\usepackage{jcappub}

\usepackage{booktabs}
\usepackage{multirow}
\usepackage{amssymb}
\usepackage[export]{adjustbox}
\usepackage{floatrow}
\newfloatcommand{capbtabbox}{table}[][3.5in]
\newfloatcommand{capbfigbox}{figure}[][2.3in]
\usepackage{blindtext}
\usepackage{amsmath}
\usepackage{booktabs}
\usepackage{aas_macros}

\def\lsim{\mathrel{\raise.3ex\hbox{$<$\kern-.75em\lower1ex\hbox{$\sim$}}}}
\def\gsim{\mathrel{\raise.3ex\hbox{$>$\kern-.75em\lower1ex\hbox{$\sim$}}}}

\begin{document}

\hspace*{110mm}{\large \tt FERMILAB-PUB-15-411-A}
\hspace*{118.3mm}{\large \tt INT-PUB-16-005}

\vskip 0.2in

\title{Is The Gamma-Ray Source \\ 3FGL J2212.5+0703 A Dark Matter Subhalo?}

\author[a,b,c,d]{Bridget Bertoni}
\emailAdd{bbertoni@stanford.edu}

\author[e,f]{and Dan Hooper}
\emailAdd{dhooper@fnal.gov}

\author[g,h]{Tim Linden}
\emailAdd{linden.70@osu.edu}

\affiliation[a]{Institute for Nuclear Theory, University of Washington,Seattle, WA 98195}      
\affiliation[b]{Stanford Institute for Theoretical Physics, Department of Physics, Stanford University, Stanford, CA 94305}
\affiliation[c]{Kavli Institute for Particle Astrophysics and Cosmology, Stanford University, Stanford, CA 94305}
\affiliation[d]{Department of Physics, University of Washington, Seattle, WA 98195}
\affiliation[e]{Fermi National Accelerator Laboratory, Center for Particle
Astrophysics, Batavia, IL 60510}
\affiliation[f]{University of Chicago, Department of Astronomy and Astrophysics, Chicago, IL 60637}
\affiliation[g]{Ohio State University, Center for Cosmology and AstroParticle Physcis (CCAPP), Columbus, OH  43210}
\affiliation[h]{University of Chicago, Kavli Institute for Cosmological Physics, Chicago, IL 60637}

\abstract{In a previous paper, we pointed out that the gamma-ray source 3FGL J2212.5+0703 shows evidence of being spatially extended. If a gamma-ray source without detectable emission at other wavelengths were unambiguously determined to be spatially extended, it could not be explained by known astrophysics, and would constitute a smoking gun for dark matter particles annihilating in a nearby subhalo. With this prospect in mind, we scrutinize the gamma-ray emission from this source, finding that it prefers a spatially extended profile over that of a single point-like source with 5.1$\sigma$ statistical significance. We also use a large sample of active galactic nuclei and other known gamma-rays sources as a control group, confirming, as expected, that statistically significant extension is rare among such objects. We argue that the most likely (non-dark matter) explanation for this apparent extension is a pair of bright gamma-ray sources that serendipitously lie very close to each other, and estimate that there is a chance probability of $\sim$2\% that such a pair would exist somewhere on the sky. In the case of 3FGL J2212.5+0703, a model with a second gamma-ray point source at the location of a known BZCAT/CRATES radio source yields fits that are comparable in quality to those obtained for a single extended source. If 3FGL J2212.5+0703 is a dark matter subhalo, it would imply that dark matter particles have a mass of $\sim$18-33 GeV and an annihilation cross section on the order of $\sigma v \sim 10^{-26}$ cm$^3$/s (for the representative case of annihilations to $b\bar{b}$), similar to the values required to generate the Galactic Center gamma-ray excess.}

\maketitle

\section{Introduction}

A wide range of strategies are being employed in an effort to observe dark matter's non-gravitational interactions.  These include experiments designed to detect the elastic scattering of dark matter with nuclei, searches for the annihilation or decay products of dark matter, and efforts to produce dark matter in accelerators. These approaches are in many cases complementary, being sensitive to different classes of dark matter candidates, and subject to different underlying limitations and uncertainties. 

Among the dark matter searches that employ gamma-ray telescopes, many different observational targets have been pursued. Among the most promising are the Galactic Center and dwarf spheroidal galaxies. While the Galactic Center is predicted to be the single brightest source of dark matter annihilation products, the astrophysical gamma-ray backgrounds from this direction of the sky are significant, and not necessarily well understood. In contrast, while the backgrounds from dwarf galaxies are expected to be quite low, the signal from annihilating dark matter in these systems is also predicted to be very faint -- orders of magnitude below that from the Galactic Center. 

The strengths and weaknesses of these approaches are well illustrated by the current status of the GeV gamma-ray excess observed from the region surrounding the Galactic Center~\cite{TheFermi-LAT:2015kwa,Daylan:2014rsa,Calore:2014xka,Abazajian:2014fta,Hooper:2013rwa,Gordon:2013vta,Abazajian:2012pn,Hooper:2011ti,Hooper:2010mq,Goodenough:2009gk}. Despite the fact that this excess has been detected at high statistical significance, and has been shown to exhibit a spectrum and morphology that is in good agreement with the predictions of annihilating dark matter~\cite{Daylan:2014rsa,Calore:2014xka}, it has been difficult to rule out other explanations for this signal, such as a large population of unresolved gamma-ray sources~\cite{Cholis:2014lta,Lee:2015fea,Bartels:2015aea,Petrovic:2014xra,Hooper:2013nhl} or a series of cosmic ray outbursts~\cite{Petrovic:2014uda,Cholis:2015dea,Carlson:2014cwa}. Constraints from dwarf spheroidal galaxies~\cite{Drlica-Wagner:2015xua,Geringer-Sameth:2014qqa} have not been able to confirm or rule out a dark matter interpretation of this signal, and the only published gamma-ray excess from any dwarf galaxy is that from the newly discovered Reticulum II~\cite{Bechtol:2015cbp,Koposov:2015cua}, with a modest statistical significance of 2.4--3.2$ \sigma$~\cite{Geringer-Sameth:2015lua,Drlica-Wagner:2015xua,Hooper:2015ula}. Without significantly more data, or the discovery of other nearby dwarf galaxies~\cite{Drlica-Wagner:2015ufc}, these strategies do not appear likely to resolve this situation in the near future. 
      
In this paper, we consider nearby dark matter subhalos as an alternative class of targets for gamma-ray telescopes.  Within the standard paradigm of cold, collisionless dark matter, the dark matter halos that host galaxies are predicted to contain very large numbers of smaller subhalos. Although the largest members of this subhalo population may host dwarf galaxies, a much larger number of subhalos are not massive enough to retain gas or form stars, and are thus effectively invisible at most wavelengths. It has been appreciated for some time, however, that nearby subhalos could be an attractive target for indirect dark matter searches using gamma-ray telescopes~\cite{Kuhlen:2008aw,Pieri:2007ir,Baltz:2008wd,Springel:2008by,Springel:2008zz,Koushiappas:2003bn,Tasitsiomi:2002vh}. For dark matter particles with an annihilation cross section near the current upper limits, Fermi would be expected to detect several such objects as high significance gamma-ray sources~\cite{Bertoni:2015mla,Berlin:2013dva,Belikov:2011pu,Buckley:2010vg,Ackermann:2012nb,Zechlin:2011wa,Mirabal:2012em,Mirabal:2010ny,Zechlin:2012by,Schoonenberg:2016aml}. 

In a recent study~\cite{Bertoni:2015mla}, we examined the Third Fermi Gamma-Ray Source Catalog (the 3FGL)~\cite{TheFermi-LAT:2015hja} in an effort to identify dark matter subhalo candidates, and to use the population of such sources to constrain the dark matter annihilation cross section. In that study, we identified 24 high latitude ($|b|>20^{\circ}$), unassociated gamma-ray sources that are very bright ($F_{\gamma}>7 \times 10^{-10}$ cm$^{-2}$ s$^{-1}$), show no evidence of variability, and that exhibit a spectral shape compatible with annihilating dark matter. While encouraging, these features alone are not particularly uncommon, and many of these sources are likely to be astrophysical objects, such as gamma-ray pulsars. Of these 24 subhalo candidates, however, the source 3FGL J2212.5+0703 stands out as particularly interesting. Unlike the other sources under consideration, the distribution of photons associated with 3FGL J2212.5+0703 does not appear to be consistent with a point source, but instead favors a spatially extended origin, with an angular radius of $\sim$$0.2^{\circ}$~\cite{Bertoni:2015mla}. If this were to be unambiguously confirmed, it would allow us to rule out all plausible astrophysical interpretations. The robust detection of an extended gamma-ray source without observable counterparts at other wavelengths would constitute a ``smoking gun'' for annihilating dark matter.

In this study, we revisit the gamma-ray source 3FGL J2212.5+0703, attempting to more precisely characterize its spatial morphology and to test the evidence in favor of its spatial extension. We confirm that the emission attributed to this source does not appear to originate from a single point source, but instead favors a spatially extended profile with high statistical significance. We acknowledge, however, that the appearance of spatial extension could potentially arise from a group of two or more bright gamma-ray sources that happen to lie within a fraction of a degree of each other on the sky. We estimate that there is an approximately two percent probability that such a source pair exists, and find that the Fermi data cannot presently exclude this possibility.

\section{Why Spatial Extension would be a ``Smoking Gun'' for Annihilating Dark Matter}

In the introduction of this paper, we asserted that a robust detection of an extended gamma-ray source without observable counterparts at other wavelengths would constitute a ``smoking gun'' for annihilating dark matter. We recognize, however, that this conclusion might strike some readers as controversial.  In this section, we discuss this issue further, and argue that any spatially extended gamma-ray source must also produce easily observable emission at other wavelengths.

In order for an astrophysical object or system to be spatially extended at a level detectable by Fermi, it must be of a physical size of at least $\sim \tan 0.1^{\circ} \times d \sim 4 \times 10^4 \,{\rm AU} \times (d/100 \,{\rm pc})$, where $d$ is the distance to the source. Such a source can clearly not be a pulsar or any other variety of compact object. Instead, spatially extended astrophysical gamma-ray sources generate their gamma-ray emission through the interactions of cosmic ray electrons and/or protons with a surrounding diffuse target of gas or radiation, via pion production, inverse Compton scattering, and/or Bremsstrahlung. In addition to any gamma-rays, such cosmic rays will also invariably generate radio emission via synchrotron. Furthermore, the heating of the diffuse material by the associated shock waves will generate radiation at a combination of X-ray, ultraviolet, visible and/or infrared wavelengths.  And in contrast to compact objects, the multi-wavelength emission generated in diffuse environments is not readily absorbed or significantly beamed, making the detectability of these accompanying signals all but inevitable. It is for these reasons that an unambiguously extended gamma-ray source, without counterparts at other wavelengths, would constitute a ``smoking gun'' for annihilating dark matter.

Moving from theoretical to empirical arguments, we point out that bright multi-wavelength emission has been detected from every extended gamma-ray source observed by Fermi (excluding 3FGL J2212.5+0703). The astrophysical objects known to produce spatially extended gamma-ray emission include pulsar wind nebulae (PWN), supernova remnants (SNR), molecular clouds, galaxy clusters, and nearby galaxies. More specifically, the Fermi Collaboration has reported spatial extension from 25 3FGL sources~\cite{TheFermi-LAT:2015hja} (see Table~\ref{extassociated}), 21 of which are associated with known SNR or PWN.\footnote{The source 3FGL J1615.3-5146e is spatially coincident with a massive star cluster, which itself contains several possible gamma-ray sources, including SNRs detected at X-ray wavelengths (by both Suzaku and XMM-Newton)~\cite{Sakai:2011sv,Acero:2013xta}. Although five pulsars have also been detected within this system, none appear to be luminous enough to power a PWN capable of producing the observed gamma-ray emission~\cite{Rowell:2009zza}. Despite the unclear origin of the gamma-ray emission from this source, it is very bright in radio, X-ray, and other wavelengths.}
 The other four of these sources are the star-forming region (SFR) Cygnus X, the lobes of the radio galaxy Centaurus A, and the satellite galaxies known as the Large and Small Magellanic Clouds (LMC, SMC). All but three of these sources are located near the Galactic Plane ($|b|<10^{\circ}$); only the LMC, SMC and Centaurus A are located at higher latitudes. We also point out that all 25 of these sources are quite bright in gamma-rays, as is necessary for any source from which Fermi could detect spatial extension.

\begin{table}
\begin{tabular}{|c|c|c|c|c|c|c|c|}
\hline
3FGL Name   &     $F_{\gamma} ({\rm cm}^{-2} \,{\rm s}^{-1})$ & $l$ & $b$ & Source Type & Source Name(s)  \\
\hline
J0852.7-4631e &	$1.30 \times 10^{-8}$ &	266.49$^{\circ}$	& -1.23$^{\circ}$	 & SNR	& 	Vela Jr, RX J0852.0-4622 \\
J0822.6-4250e	& $8.30 \times 10^{-9}$	&	260.32$^{\circ}$&	-3.28$^{\circ}$ &	SNR	 &	Puppis A \\	
J1713.5-3945e	& $4.54 \times 10^{-9}$ &		347.34$^{\circ}$ &	-0.47$^{\circ}$ &	SNR	& 	RX J1713.7-3946	\\ 
J1801.3-2326e	& $5.35\times 10^{-8}$ &		6.53$^{\circ}$	& -0.25$^{\circ}$ &	SNR	 &	W28	\\
J1805.6-2136e	& $2.42\times 10^{-8}$ &		8.60$^{\circ}$	&-0.21$^{\circ}$ &	SNR	 &	W30	 \\
\hline
J1855.9+0121e	&$7.07\times 10^{-8}$ &	34.65$^{\circ}$	&-0.39$^{\circ}$	&SNR&	 	W44, 3C392 \\
J1923.2+1408e	&$3.96\times 10^{-8}$ &	49.12$^{\circ}$	&-0.46$^{\circ}$	&SNR&	 	W51C\\	
J0617.2+2234e	&$6.33\times 10^{-8}$ &	189.05$^{\circ}$	&3.03$^{\circ}$	&SNR	& 	IC 443, 3C157 \\ 
J0540.3+2756e	&$7.05\times 10^{-9}$ &	180.24$^{\circ}$	&-1.50$^{\circ}$	&SNR	 &	S 147\\	 
J2051.0+3040e	&$1.02\times 10^{-8}$ &	73.98$^{\circ}$	&-8.56$^{\circ}$	&SNR	 &	Cygnus Loop\\	 
\hline
J2021.0+4031e	&$6.90\times 10^{-10}$ &	78.24$^{\circ}$&	2.20$^{\circ}$	&SNR	 &	$\gamma$-Cygni, VER J2019+407	\\
J2045.2+5026e	&$1.07\times 10^{-8}$ &	88.75$^{\circ}$&	4.67$^{\circ}$	&SNR	 &	HB 21	 \\
J1615.3-5146e	&$9.26\times 10^{-8}$ &	331.66$^{\circ}$	&-0.66$^{\circ}$& SNR/PWN	&	HESS J1614-518\\	
 J1303.0-6312e	&$1.57	\times 10^{-9}$ &	304.23$^{\circ}$	&-0.36$^{\circ}$	&PWN	& 	HESS J1303-631\\
 J1514.0-5915e	&$2.90	\times 10^{-9}$ &	320.27$^{\circ}$	&-1.27$^{\circ}$	&PWN	 &	MSH 15-52	\\
 \hline
 J1616.2-5054e	&$1.39	\times 10^{-8}$ &	332.37$^{\circ}$	&-0.13$^{\circ}$	&PWN	 &	HESS J1616-508 \\
 J1633.0-4746e	&$2.26\times 10^{-8}$ &		336.52$^{\circ}$&	0.12$^{\circ}$&	PWN	 &	HESS J1632-478\\	
 J0833.1-4511e	  &$   	1.83\times 10^{-8}$ &	263.33$^{\circ}$&	-3.10$^{\circ}$&	PWN	& 	Vela X	\\
 J1824.5-1351e	&$7.84	\times 10^{-9}$ &	17.57$^{\circ}$&	-0.45$^{\circ}$	&PWN	& 	HESS J1825-137\\	
 J1836.5-0655e	&$9.65	\times 10^{-9}$ &	25.08$^{\circ}$&	0.14$^{\circ}$	&PWN	 &	HESS J1837-069\\	
 \hline
 J1840.9-0532e	&$1.18	\times 10^{-8}$ &	26.80$^{\circ}$	&-0.20$^{\circ}$&	PWN	 &	HESS J1841-055\\	
  J0059.0-7242e&$	3.50\times 10^{-9}$ &	302.14$^{\circ}$&	-44.42$^{\circ}$&	Galaxy	& 	SMC	 \\
 J0526.6-6825e	 &$2.02\times 10^{-8}$ &		278.84$^{\circ}$	&-32.85$^{\circ}$	&Galaxy	 &	LMC	 \\
 J1324.0-4330e &$	3.39	\times 10^{-9}$ &		309.17$^{\circ}$&	18.98$^{\circ}$&	Radio Gal.	 &	Centaurus A (lobes)	 \\
  J2028.6+4110e&$	5.80	\times 10^{-8}$ &		79.60$^{\circ}$&	1.40$^{\circ}$&	SFR	 &	Cygnus X Cocoon	\\
\hline
\end{tabular}
\caption{The 25 gamma-ray sources identified as being spatially extended in the 3FGL catalog~\cite{TheFermi-LAT:2015hja}. These sources are each observed to produce bright emission at a combination of radio, X-ray, TeV-scale, and other wavelengths.}
\label{extassociated}
\end{table}

For the purposes of this study, the most important feature of this collection of sources is that they are {\it all} very bright and easily detectable at other wavelengths. For example, each of the SNRs listed in Table~\ref{extassociated} emits very bright radio emission, with flux densities (at 1 GHz) in the range of 30 to 320 Jy~\cite{Green:2014cea}.\footnote{A Jansky (Jy) is a unit of radio flux density, equivalent to $10^{-26}$ watts per square meter, per hertz.}  
%
Similarly, each of the PWN in this list have been detected by ground-based telescopes at TeV-scale energies, as have most of the SNR, as well as the Cygnus X Cocoon~\cite{TheFermi-LAT:2015hja}. The LMC and SMC have each been detected in X-ray, ultra-violet, visible, infrared, and radio wavelengths. Furthermore, all 25 of these sources have been detected at X-ray wavelengths~\cite{chandracatalog,Dubner:2013bpa,Kargaltsev:2007kf,2013ApJ...771...91B,2012ApJ...752..135K,2013MNRAS.436..968L,2001AAS...198.4002L,Kargaltsev:2008hf,Sakai:2011sv,Acero:2013xta,Nobukawa:2015jva,Mizuno:2015jua}. If 3FGL J2212.5+0703 is, in fact, spatially extended, its lack of bright multi-wavelength counterparts makes it very unlike any of the previously detected extended gamma-ray sources.

\section{Dark Matter Subhalo Candidates}
\label{candidates}

In early 2015, the Fermi Collaboration released its Third Gamma-Ray Source Catalog, the 3FGL~\cite{TheFermi-LAT:2015hja}. This catalog contains an impressive 3033 sources, of which 992 had not been associated with emission observed at other wavelengths. It is all but certain that the overwhelming majority of these 992 unassociated sources are not dark matter subhalos, but are instead a collection of more conventional astrophysical objects, such as radio-faint pulsars or distant active galactic nuclei (AGN).\footnote{Many unassociated 3FGL sources are classified as unassociated due to a large location error and not due to a lack of potential associations.  Additionally, a large location error complicates gamma-ray pulsation searches, making pulsar associations more difficult.} In contrast to the roughly isotropic distribution predicted for detectable dark matter subhalos, most galactic gamma-ray sources (including most pulsars and supernova remnants) are distributed preferentially around the Galactic Plane. For this reason, we limit our subhalo search to those 380 unassociated sources in the 3FGL that are located at high galactic latitudes, $|b|>20^{\circ}$. Furthermore, as it is very difficult to detect spatial extension from any but the brightest gamma-ray sources~\cite{Lande:2012xn}, we limit our study to those with a gamma-ray flux greater than $F_{\gamma} > 10^{-9}$ cm$^{-2}$ s$^{-1}$ (integrated above 1 GeV). We also remove the source 3FGL J0536.4-3347 from our list of dark matter subhalo candidates, as it is flagged in the 3FGL as residing ``on top of an interstellar gas clump or small-scale defect in the model of diffuse emission.'' The 3FGL contains 17 unassociated sources that meet these requirements

Over the past year or so, evidence has been presented in favor of astrophysical interpretations for five of these sources. In particular, a millisecond pulsar (MSP) was recently detected with the Parkes radio telescope in the direction of 3FGL J1946.4-5403~\cite{camilo}. Similarly, emission at optical and X-ray wavelengths in association with 3FGL J1544.6-1125 was recently reported, suggesting a likely classification for this source as a MSP binary~\cite{Bogdanov:2015xda}. Additionally, the authors of Ref.~\cite{2015ApJS..217....4S} have associated 3FGL J2103.7-1113 with a radio source, interpreting it as a likely AGN, and Ref.~\cite{Dai:2016det} has reported the detection of X-ray emission from the directions of 3FGL J1120.6+0713 and 3FGL J2103.7-1113, providing support for AGN origins. And lastly, gamma-ray pulsations were recently detected from 3FGL J1744.1-7619~\cite{camilo}. In light of this new information, we exclude these five sources from our primary list of dark matter subhalo candidates, focusing our attention on the remaining 12 sources.\footnote{We thank Elizabeth Ferrara, Fernando Camilo, Paul Ray, and Frank Schninzel for bringing these recent associations to our attention.} Notably, despite being the target of multiple observations by the Arecibo telescope, 3FGL J2212.5+0703 has not yielded any detectable radio pulsations~\cite{Cromartie:2016fzm}.

\begin{table}
\begin{tabular}{|c|c|c|c|c|c|c|c|cl}
\hline
3FGL Name   &    $F_{\gamma} \,({\rm cm}^{-2} \,{\rm s}^{-1})$ & $l$ & $b$ &  $m_{\chi}$~(GeV) & Recent Association  \\
\hline
  J0318.1+0252 &$1.23\times10^{-9}$	&178.44$^{\circ}$&	-43.64$^{\circ}$& 26.0$-$42.1   &--\\
 J0523.3-2528 & $1.77\times10^{-9}$	&228.20$^{\circ}$&	$-$29.83$^{\circ}$&34.6$-$48.6&--\\ 
 J0953.7-1510 & $1.25\times10^{-9}$	&251.94$^{\circ}$	&29.61$^{\circ}$&27.1$-$43.5&--\\
 J1050.4+0435 & $1.06\times10^{-9}$&	245.55$^{\circ}$	&53.41$^{\circ}$&--&--\\ 
 \hline
 J1119.9-2204  & $2.70\times10^{-9}$	&276.47$^{\circ}$	&36.06$^{\circ}$&18.6$-$25.3&--\\
 J1120.6+0713 & $1.10\times10^{-9}$&251.53$^{\circ}$&	60.69$^{\circ}$&32.1$-$55.1& AGN\,(?)~\cite{Dai:2016det}\\
 J1225.9+2953  & $1.42\times10^{-9}$	&185.1$^{\circ}$&	83.76$^{\circ}$&40.7$-$63.7&--\\
 J1544.6-1125  & 	$1.01\times10^{-9}$&	356.21$^{\circ}$&	32.98$^{\circ}$&11.8$-$19.8& MSP~\cite{Bogdanov:2015xda}\\
 \hline
 J1548.4+1455  &$1.30\times10^{-9}$	&25.63$^{\circ}$&	47.18$^{\circ}$&--&--\\
  J1625.1-0021  & $3.57\times10^{-9}$	&13.88$^{\circ}$	&31.84$^{\circ}$&33.5$-$42.1&--\\
   J1653.6-0158  & $4.24\times10^{-9}$	&16.62$^{\circ}$	&24.92$^{\circ}$&--&--\\
    J1744.1-7619 & $3.85\times10^{-9}$	&317.10$^{\circ}$	&-22.47$^{\circ}$&26.8$-$31.8& Pulsar~\cite{camilo}\\
    \hline
J1946.4-5403 & $1.72\times10^{-9}$&	343.89$^{\circ}$	&-29.56$^{\circ}$&21.1$-$31.4& MSP~\cite{camilo}\\
 J2039.6-5618 & $2.32\times10^{-9}$	&341.23$^{\circ}$&	-37.15$^{\circ}$&33.3$-$48.0& --\\
  J2103.7-1113  & $1.09\times10^{-9}$&	37.8$^{\circ}$	&-34.42$^{\circ}$&29.7$-$51.4&AGN~\cite{2015ApJS..217....4S,Dai:2016det}\\
    J2112.5-3044& $3.26\times10^{-9}$	&14.90$^{\circ}$&	-42.45$^{\circ}$&40.2$-$53.5&--\\
    J2212.5+0703 &  $1.24\times10^{-9}$	&68.74$^{\circ}$&	-38.56$^{\circ}$& 18.4$-$32.7& --\\ 
\hline
\end{tabular}
\caption{The gamma-ray flux (above 1 GeV), galactic longitude, and galactic latitude of the 17 sources classified as unassociated in the 3FGL with a flux above $10^{-9}$ cm$^{-2}$ s$^{-1}$ and located at $|b|>20^{\circ}$. In the rightmost column, we make note of the five sources that have been recently associated with emission at other wavelengths, or had gamma-ray pulsations identified. Also listed are the range of dark matter masses found to provide a good fit to the measured gamma-ray spectrum of each source (for the representative case of annihilations to $b\bar{b}$). For sources with no entry in this column, no value of the dark matter mass was found to provide a good fit.}
\label{candidatesummary}
\end{table}

In Table~\ref{candidatesummary}, we list the gamma-ray flux, galactic longitude, and galactic latitude for each of the 12 unassociated gamma-ray sources passing the aforementioned latitude and flux requirements, as well as for the five recently associated sources described in the previous paragraph. For each source that is well-fit ($2 \Delta \ln \mathcal{L} < 1.3$ per degree-of-freedom) by annihilating dark matter for some choice of the mass, we also provide in this table the range of dark matter masses that lie within $2 \Delta \ln \mathcal{L} < 4$ of the best-fit value (for the representative case of annihilations to $b\bar{b}$). For sources with no entry in this column, no value of the dark matter mass was found to provide a good fit.  After excluding those sources with recent associations, we find that nine of these candidates sources are well fit by some range of dark matter masses, approximately eight of which prefer values near the range favored by the Galactic Center gamma-ray excess~\cite{Daylan:2014rsa,Calore:2014xka}.


To study the spectrum and morphology of this selection of 3FGL
sources, we make use of approximately 7 years of Fermi-LAT
data,\footnote{MET range 239557417 - 462376143} utilizing Pass 8
photons in the energy range of 0.1 to 100 GeV. We have applied
standard analysis cuts, excluding events arriving at a zenith angle
greater than 90$^\circ$, as well as any event that does not pass
``Source'' photon selection criteria. We additionally exclude events
recorded while the instrument was not in science survey mode, when the
instrumental rocking angle was $>$52$^\circ$, or when the instrument
was passing through the South Atlantic Anomaly. For each source, we
consider photons observed within a 14$^\circ\times$14$^\circ$ box centered
at the source's location and divide these
photons into 280$\times$280 angular bins and 15 evenly spaced
logarithmic energy bins. To maximize our sensitivity to extended
source emission profiles, we utilize the Pass 8 photon point spread function (PSF) event
classes and analyze events in each PSF class independently. We utilize
the {\tt P8R2\_SOURCE\_V6} instrumental response functions, employing the
latest model for diffuse Galactic gamma-ray emission
({\tt gll\_iem\_v06.fits}), the latest isotropic emission for the Source
photon data selection ({\tt iso\_P8R2\_SOURCE\_V6\_PSF0\_v06.txt}) and include
all 3FGL sources which lie within 10$^\circ$ of our region of
interest, utilizing default Fermi-LAT prescriptions to determine which
sources are given freedom to float in our fits to the gamma-ray data. We fit our model to the Fermi-LAT data independently in each energy bin, and do not impose any parameterization on a source's spectral shape. To calculate the best-fit
flux from each source in a given energy bin, we use
the Fermi-LAT {\tt pyLikelihood} code, utilizing the {\tt MINUIT} algorithm. 
For each subhalo candidate, we additionally compute the full likelihood
profile in each energy bin and for each PSF class, in order to
accurately determine the source spectrum.


\begin{figure}
\begin{center}
\includegraphics[width=0.99\linewidth]{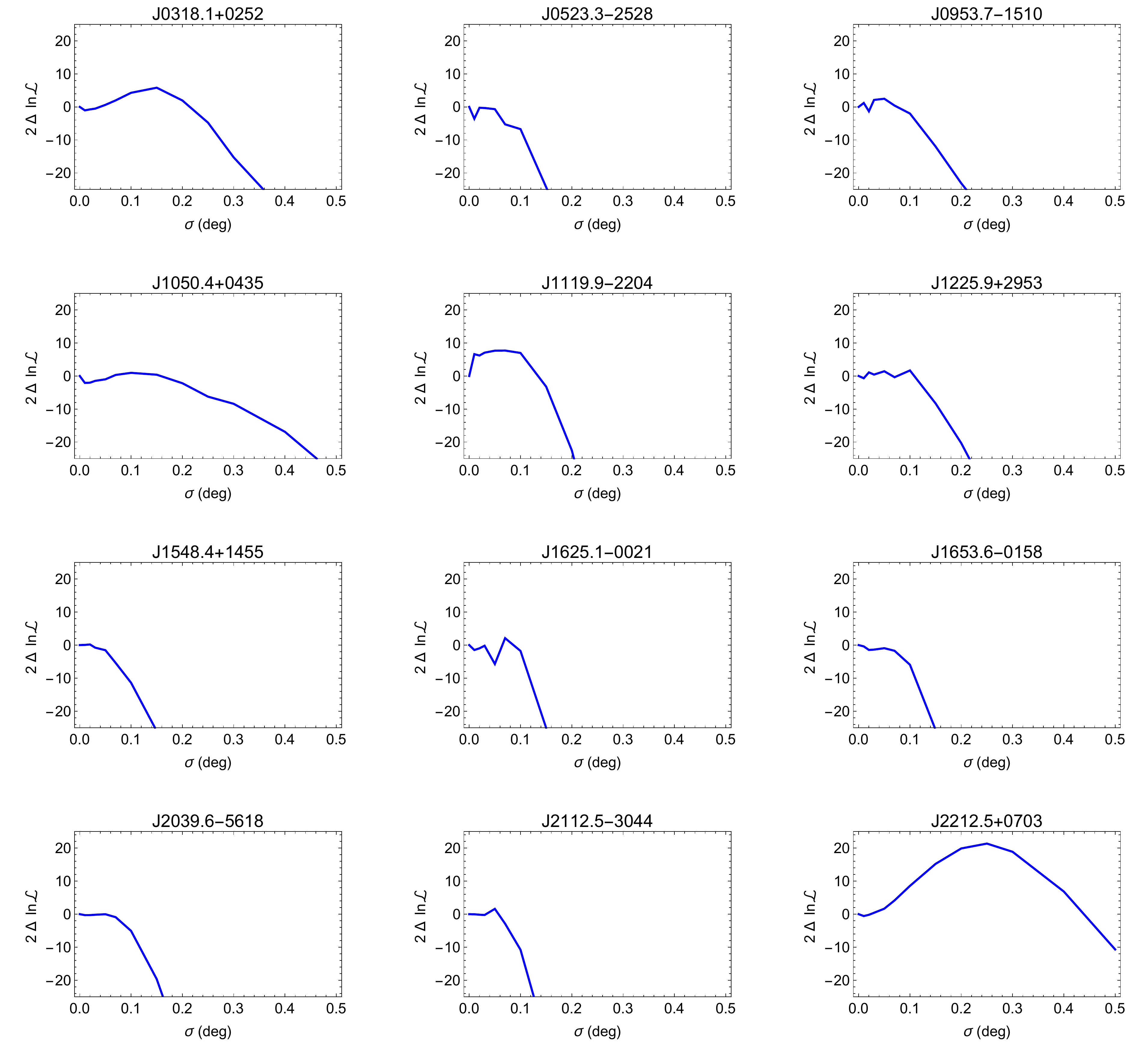}
\end{center}
\caption{The change to the log-likelihood as a function of (spherically symmetric) spatial extension, for the 12 subhalo candidate sources considered in our study. The source 3FGL J2212.5+0703 shows significant evidence of spatial extension.}
\label{extunass1}
\end{figure}

For each subhalo candidate source, we test for spatial extension by replacing the point-source template with an extended template, varying the width as a free parameter. For the profile of the extension, we adopt a distribution corresponding to a Navarro-Frenk-White (NFW) dark matter profile, truncated by the effects of tidal stripping such that only the innermost 0.5\% of the subhalo's mass remains intact~\cite{Berlin:2013dva,Springel:2008cc}. We calculate the angular and energy distribution of photons as follows:
\begin{align}
\label{flux1}
\Phi (E_{\gamma}, \theta) = \frac{1}{8 \pi  m_{\chi}^2}  \langle \sigma v \rangle \frac{\mathrm{d}N_{\gamma}}{\mathrm{d}E_\gamma}  \,  \int_{\text{l.o.s.}} \rho^2 [ r(d,l,\theta)] \,\mathrm{d}l, 
\end{align}
where $m_{\chi}$ is the mass of the dark matter particle, $\langle \sigma v \rangle$ is the annihilation cross section, and $dN_{\gamma}/dE_{\gamma}$ is the gamma-ray spectrum produced per annihilation, which we calculate using PYTHIA 8~\cite{pythia}. The integral of the density squared is performed over the line-of-sight, $d$ is the distance to the center of the subhalo, $\theta$ is the angle to the center of the subhalo, and $r(\theta,d,l)=\sqrt{d^2+l^2-2dl\cos{\theta}}$. The NFW density profile itself (prior to tidal truncation) is given by~\cite{Navarro:1995iw,Navarro:1996gj}:
\begin{equation}
\rho(r) = \frac{\rho_0}{(r/R_s) [1+(r/R_s)]^2}.
\end{equation}
To determine the subhalo's scale radius, $R_s$, we adopt the mass-concentration relationship described in Ref.~\cite{Sanchez-Conde:2013yxa}. We define the width by the parameter $\sigma$, which is the angular radius that contains 68\% of the photons from the source. 

In Fig.~\ref{extunass1}, we plot the change to the log-likelihood, for each subhalo candidate source when the point-like template is replaced with that of an extended source.  Of these 12 high-latitude ($|b|>20^{\circ}$), bright ($F_{\gamma} > 10^{-9}$ cm$^{-2}$ s$^{-1}$), unassociated sources, the most significant evidence for extension is from 3FGL J2212.5+0703, for which $2 \ln {\mathcal L}$ increases by 21.4 when the point-like template is replaced by a spherically symmetric, tidally stripped NFW profile with a width of $\sigma=0.25^{\circ}$. In Table~\ref{candidates}, we present the results of our test for spatial extension for these 12 dark matter subhalo candidates. For the seven sources that prefer an extended profile (at a level of $2\Delta \ln {\mathcal L}$~$>$~1.0), we provide the best-fit value for their extension parameter, $\sigma$. For each of these 12 sources, we provide the $2\sigma$ upper limit on the degree of spatial extension (corresponding to the value of $\sigma$ for which $2\ln {\mathcal L}_{\rm point} - 2\ln {\mathcal L}_{\rm ext}= 4$).


\begin{table}
\begin{tabular}{|c|c|c|c|c|c|}
\hline
 \multicolumn{1}{|c|}{Source Name (3FGL)} &  $\sigma$ & $2 \Delta \ln {\mathcal L}$   \\
 \hline
J2212.5+0703 & 0.25$^{\circ}$ ($<0.31$$^{\circ}$)  & 21.4 \\ 
J1119.9-2204  & $0.07^{\circ}$ ($<0.12$$^{\circ}$)  & 7.7 \\
 J0318.1+0252 &0.15$^{\circ}$ ($<0.20$$^{\circ}$)  & 5.8 \\ 
 J0953.7-1510 & $0.05^{\circ}$ ($<0.09$$^{\circ}$) & 2.5\\
 J1625.1-0021  & $0.07^{\circ}$ ($<0.10^{\circ}$) & 2.1 \\
  J1225.9+2953  & $0.10^{\circ}$ ($<0.12$$^{\circ}$) & 1.7 \\
J2112.5-3044& $0.05^{\circ}$ ($<0.07$$^{\circ}$)& 1.6 \\
 J0523.3-2528 & $<0.06$$^{\circ}$ & -- \\
 J1050.4+0435 & $<0.21^{\circ}$ & -- \\
 J1548.4+1455  & $<0.06^{\circ}$ & -- \\
 J1653.6-0158  & $<0.09^{\circ}$ & -- \\
 J2039.6-5618 & $<0.09^{\circ}$ & -- \\
\hline
\end{tabular}
\caption{The results of our test for spatial extension for the 12 bright ($F_{\gamma} > 10^{-9}$ cm$^{-2}$ s$^{-1}$) and high-latitude ($|b|>20^{\circ}$) dark matter subhalo candidates. For 3FGL J2212.5+0703, we find significant evidence in favor of spatial extension. For the next six sources listed, the fit modestly prefers a spatially extended distribution (at a level of $2\Delta \ln {\mathcal L}$~$>$~1.0). For these seven sources, we provide the best-fit value for their extension parameter, $\sigma$. The other five sources show no significant preference for any spatial extension. For each source, we provide the $2\sigma$ upper limit on the degree of spatial extension (corresponding to the value of $\sigma$ for which $2\ln {\mathcal L}_{\rm point} - 2\ln {\mathcal L}_{\rm ext}= 4$).}
\label{candidates}
\end{table}

\begin{figure}
\begin{center}
\includegraphics[width=0.99\linewidth]{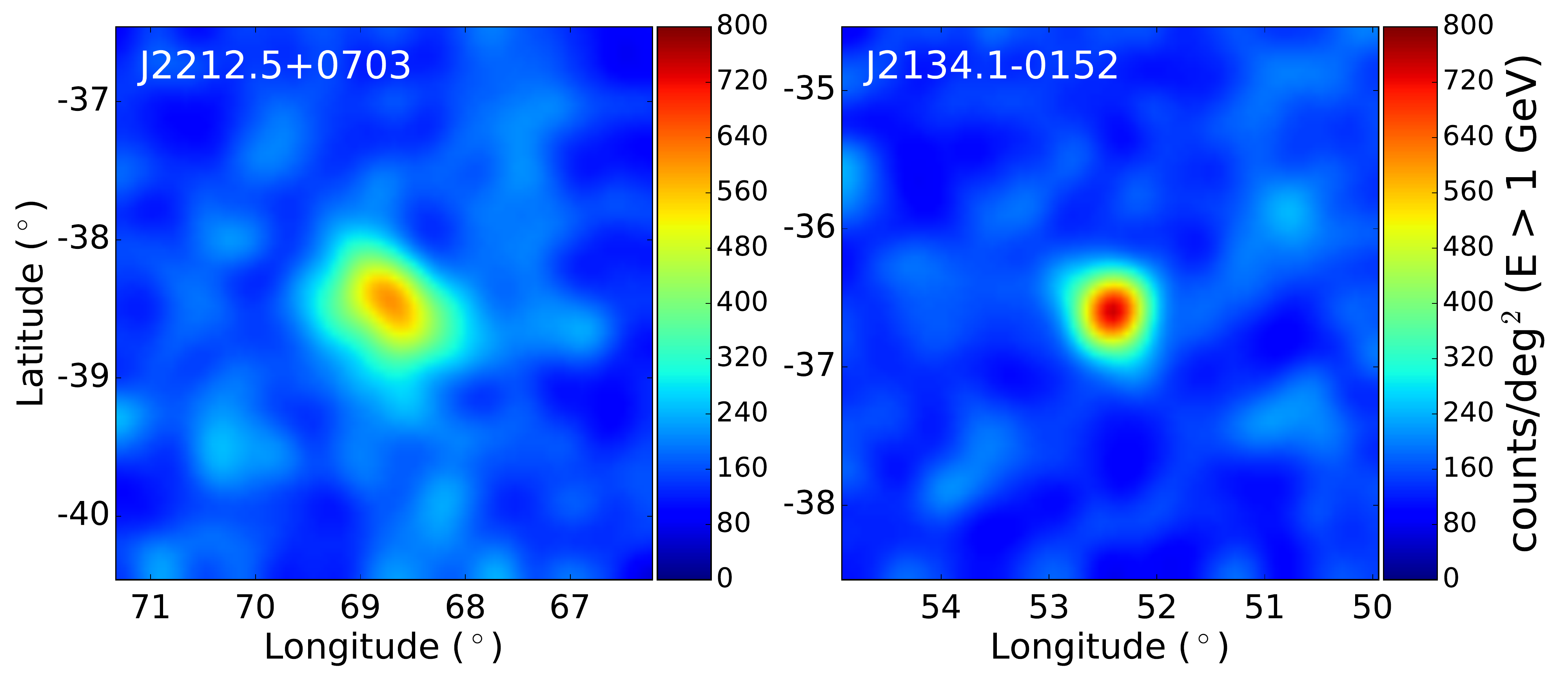}
\end{center}
\caption{Residual maps of the regions surrounding the subhalo candidate 3FGL J2212.5+0703 (left frame) and the known blazar 3FGL J2134.1-0152 (right frame). These maps display the photon flux per square degree (above 1 GeV) and have been smoothed with a 0.15$^{\circ}$ Gaussian. Whereas the source in the left frame shows significant evidence of spatial extension, the source in the right frame is consistent with point-like emission.}
\label{maps}
\end{figure}

To further investigate the morphology of 3FGL J2212.5+0703, we show in the left frame of Fig.~\ref{maps} a map of the gamma-ray residuals in the region surrounding this source. This map represents the total photon counts (above 1 GeV), after subtracting the best-fit background model (including the galactic diffuse, isotropic, and nearby point-source models). The maps has been smoothed with a Gaussian function with a smoothing length of $0.15^{\circ}$. This can be compared directly to the map shown in the right frame of this figure, which is of the point-like source 3FGL J2134.1-0152, a known blazar. The spatial extension of 3FGL J2212.5+0703 can be seen by eye when the residual maps of these two sources are compared.

Thus far, we have adopted spherically symmetric templates for the emission from 3FGL J2212.5+0703 and the other dark matter subhalo candidates. In contrast, numerical simulations find that dark matter subhalos are predominantly triaxial~\cite{Kuhlen:2007ku,Vera-Ciro:2014ita}. To explore this possibility, we replaced the spherical truncated NFW template with one stretched along one axis, allowing the extension, axis ratio, and orientation to float freely. The best fit was found for a profile with an axis ratio of $\sim$3.4 and a major axis that is oriented approximately $43^{\circ}$ counterclockwise from the vertical (in the coordinates shown in Fig.~\ref{maps}). This choice of profile improves upon the spherically symmetric template at the level of $2 \Delta \ln {\mathcal L}$=11.0, while adding two additional degrees-of-freedom. This elliptically extended profile is preferred over that of a single point source by $2 \Delta \ln {\mathcal L}$=32.4, which (for three additional degrees-of-freedom) corresponds to a statistical significance of $5.1\sigma$.

\section{Systematic Uncertainties: Assessing the Robustness of 3FGL J2212.5+0703's Spatial Extension}

In this section, we will describe tests that we have performed in order to establish the probability that the spatial extension observed from 3FGL J2212.5+0703 is authentic, as opposed to being the result of problems with the diffuse emission model or confusion between multiple nearby gamma-ray sources.

\begin{figure}
\begin{center}
\includegraphics[width=1.0\linewidth]{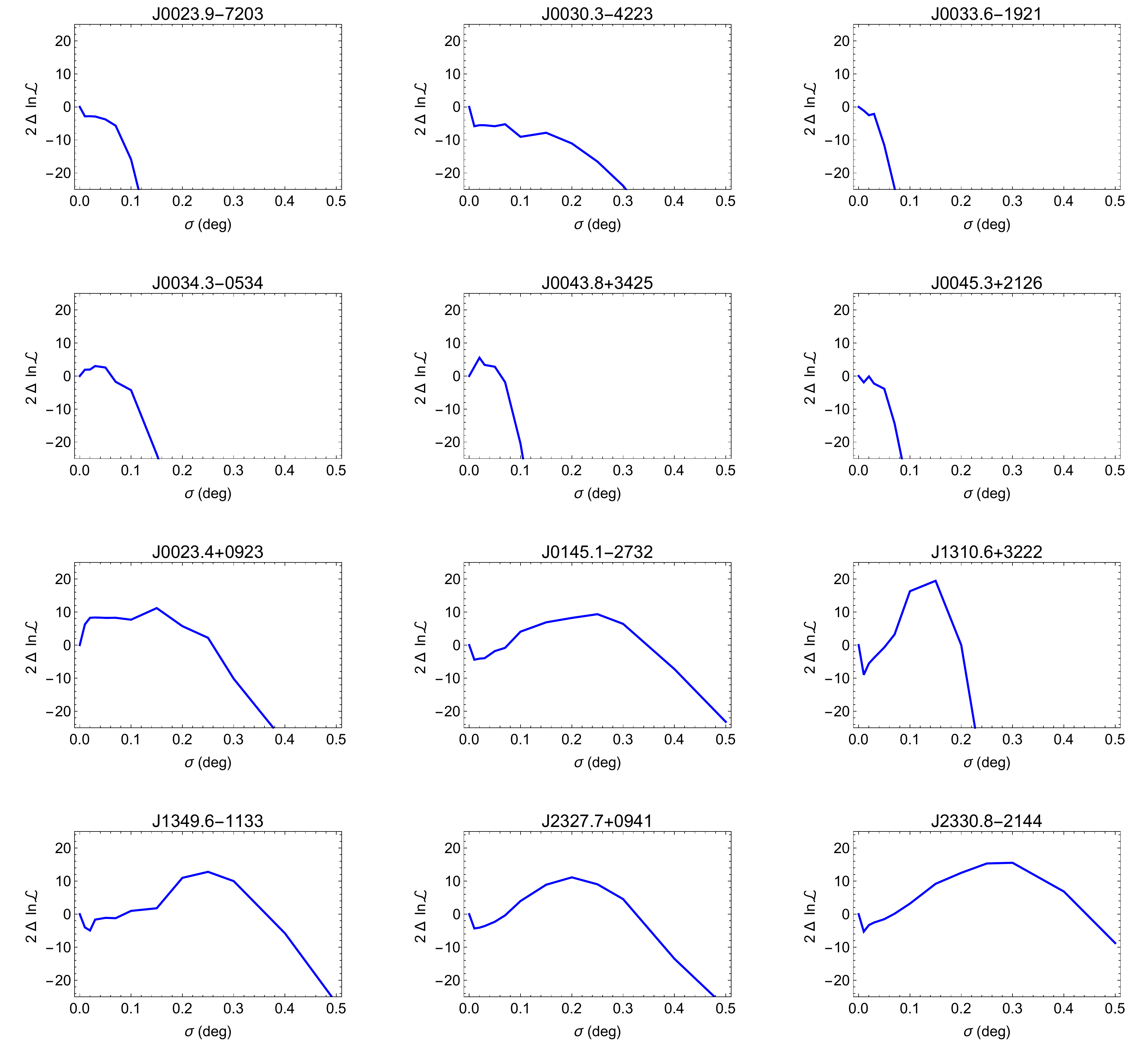}
\end{center}
\caption{The change to the log-likelihood as a function of (spherically symmetric) spatial extension, for a sample of bright, high latitude 3FGL sources that have been associated with emission observed at other wavelengths. In the upper six frames we show results for typical sources, which appear point-like, while the lower six frames describe those sources that exhibit the greatest evidence of extension.}
\label{extass}
\end{figure}

\subsection{Using Associated 3FGL Sources as a Control Group}

In addition to the unassociated sources discussed in the previous section, the 3FGL catalog contains many sources that have been associated with emission observed at other wavelengths. These sources, which are very unlikely to be dark matter subhalos, provide us with an opportunity to test our procedure for identifying spatial extension. In order to make a fair ``apples-to-apples'' comparison, we consider only those associated 3FGL sources that are located at high latitude ($|b|>20^{\circ}$) and that emit a gamma-ray flux in the same range as our 12 subhalo candidates ($10^{-9}$ cm$^{-2}$ s$^{-1} < F_{\gamma} < 4.24 \times 10^{-9}$ cm$^{-2}$ s$^{-1}$).  Of the 251 sources that meet these criteria, 228 are associated with AGN, 16 with pulsars, six with galaxies, and one with a globular cluster. 

Following the approach described in the previous section, we have tested each of these 251 sources for evidence of (spherically symmetric) spatial extension. While we found that none of these 251 sources exhibit as much evidence for extension as 3FGL J2212.5+0703, the flat-spectrum radio quasar 3FGL J1310.6+3222 does prefer extension at a slightly lower level, $2 \Delta \ln {\mathcal L}$~$\simeq 19.4$ (compared to 21.4 for 3FGL J2212.5+0703). Including this source, we found that five of these 251 sources prefer extension at the level of $2 \Delta \ln {\mathcal L}$~$> 10$.  In Fig.~\ref{extass}, we present these results for a small sub-sample of these 251 sources. In the upper six frames we show typical sources, which appear point-like, while the lower six frames depict those sources that exhibit the greatest evidence of extension.

Although this exercise demonstrates that false evidence in favor of spatial extension is somewhat rare, seemingly extended sources are about ten times more common than anticipated from statistical fluctuations alone. This rate, however, is entirely consistent with the number of high-latitude 3FGL sources that are expected to be located within $\sim0.1^{\circ}$--\,$0.3^{\circ}$ of another 3FGL source. For such a pair of sources, discrimination is expected to be difficult, allowing them to potentially appear as a single extended source. In the following subsection, we estimate the probablity that the emission attributed to 3FGL J2212.5+0703 could instead be from pair of nearby gamma-ray point sources.

\subsection{Assessing the Probability of Nearby Source Confusion}
\label{confusion}

If two or more unassociated gamma-ray point sources were located very close to each other on the sky, they could be misinterpreted as a single source, possibly with an apparent degree of spatial extension.  In this subsection, we estimate the probability of such a signal arising and assess this possibility within the context of 3FGL J2212.5+0703.

Considering a collection of $N$ gamma-ray sources, distributed randomly across the \\ $|b|> 20^{\circ}$ sky, the probability that any two will lie within an angle, $\alpha$, of each other is given by:
\begin{eqnarray}
\mathcal{P} &=& \frac{N (N-1)}{2} \,\frac{\pi \alpha^2}{4\pi \, [1-\sin 20^{\circ}]} \nonumber \\
&\approx & 2.3 \times 10^{-6} \times N (N-1) \times \bigg(\frac{\alpha}{0.2^{\circ}}\bigg)^2.
\end{eqnarray}

If we include all of Fermi's unassociated high-latitude sources ($N=380$), this calculation yields a significant probability ($\sim$33\%) that a pair of these sources will fall within $0.2^{\circ}$ of each other.  But 3FGL J2212.5+0703 is a very bright source, and most overlapping source pairs will not produce such a large flux. When we require that the combined flux of the two sources be greater than that of 3FGL J2212.5+0703 ($F_1+F_2 \ge 1.24\times 10^{-9}$ cm$^{-2}$ s$^{-1}$) and that neither source is much fainter than the other ($F_{1,2} \ge 0.2 F_{2,1}$), we obtain the following result:
\begin{eqnarray}
\mathcal{P} \approx  2.2 \times 10^{-2} \times \bigg(\frac{\alpha}{0.2^{\circ}}\bigg)^2.
\end{eqnarray}

We thus conclude that there is a small ($\sim$\,2\%), but not entirely negligible, probability that two such sources would reside close enough to each other to potentially be confused with a single bright extended source.

\subsection{Nearby Radio Sources}

If 3FGL J2212.5+0703 is actually a superposition of two nearby unassociated point-like gamma-ray sources, it is possible that one or both of these sources could be found within existing multi-wavelength catalogs. With this in mind, we have consulted the following:
\begin{itemize}
\item{The Roma-BZCAT Multi-Frequency Catalog of Blazars (BZCAT)~\citep{Massaro:2008ye}}
\item{The Combined Radio All-Sky Targeted Eight-GHz Survey (CRATES) catalog~\cite{Healey:2007by}}
\item{The Candidate Gamma-Ray Blazar Survey (CGRaBS) catalog~\cite{Healey:2007gb}}
\item{The Australia Telescope National Facility (ATNF) pulsar catalog~\cite{Hobbs:2003gk}}.
\end{itemize}

Within these four catalogs, two sources were found within $1^{\circ}$ of 3FGL J2212.5+0703.  These two radio sources (BZQ J2212+0646/CRATES J221251+064604 and \\ CRATES J221408+071128), are located at angular distances of approximately 0.3$^{\circ}$ and 0.4$^{\circ}$ away from 3FGL J2212.5+0703, respectively, and each have radio fluxes of a few hundred mJy, measured at several frequencies.

To explore the possibility that either of these BZCAT and/or CRATES sources might be responsible for the apparent extension observed from 3FGL J2212.5+0703, we re-performed the fit of the region, considering various pairs of point-source locations. In order to limit the number of new degrees-of-freedom being introduced, we adopt a power-law form for the gamma-ray spectrum from these radio sources; a well-motivated choice given the spectra observed from AGN. 

The best fit found in this exercise was for a point source at the location of BZQ J2212+0646/CRATES J221251+064604 and a second point source at or near the location of 3FGL J2212.5+0703 (the inclusion of a gamma-ray source at the location of CRATES J221408+071128 did not significantly improve the fit). If we fix the location of 3FGL J2212.5+0703 to its best-fit position as given in the 3FGL catalog, the addition of a gamma-ray point source at the location of this BZCAT/CRATES source improves the fit (over a single point source) at the level of $2 \Delta \ln {\mathcal L} = 26.4$, at the expense of introducing two new degrees-of-freedom (normalization and spectral index of the additional source). If we further allow the location of 3FGL J2212.5+0703 to float in the fit, the improvement is $2 \Delta \ln {\mathcal L} = 33.0$. Including the fact that this fit includes one more degree-of-freedom than in the case of a single elliptical extended source (for which we found $2 \Delta \ln {\mathcal L} = 32.4$), we find that these two hypotheses are approximately equally supported by the data. Thus at this time, we cannot discriminate between these two possible interpretations for 3FGL J2212.5+0703.

\section{The Implications of 3FGL J2212.5+0703 as a Dark Matter Subhalo}

In this section, we will assume that 3FGL J2212.5+0703 is in fact a subhalo of annihilating dark matter particles
and consider what this observation implies about this particular subhalo, and about the nature of dark matter itself. 

\begin{figure}
\begin{center}
\includegraphics[width=0.7\linewidth]{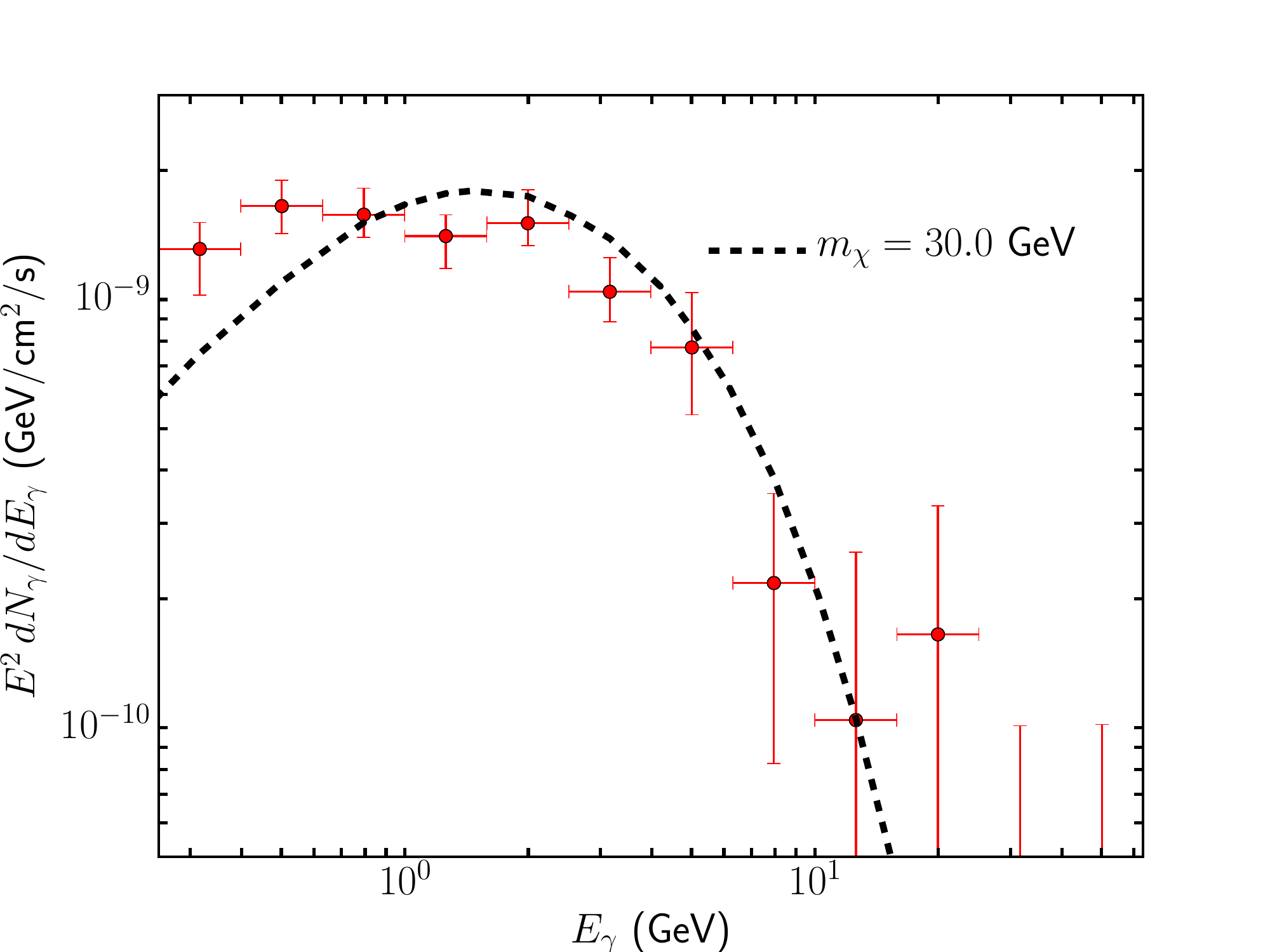}
\end{center}
\caption{The gamma-ray spectrum of 3FGL J2212.5+0703. The dashed curve denotes the spectral shape predicted from a 30 GeV dark matter particle that annihilates to $b\bar{b}$. Dark matter masses in the range of 18.4-32.7 GeV provide a good fit to the measured spectrum.}
\label{spectrum}
\end{figure}

We begin by considering the spectrum of the gamma-ray emission observed from this source, which we plot in Fig.~\ref{spectrum}. For the case of dark matter annihilating to $b\bar{b}$, masses in the range of 18.4-32.7 GeV provide a good fit at the 2$\sigma$ ($2\Delta \ln {\mathcal L}=4$) level.  For dark matter that annihilates into light quarks (gauge/Higgs bosons), lower (higher) masses can also provide a good fit to the measured spectrum. We also note that the spectrum of 3FGL J2212.5+0703 is similar to that of the Galactic Center excess~\cite{Calore:2014nla,Calore:2014xka,Daylan:2014rsa,TheFermi-LAT:2015kwa}.


Numerical simulations including Aquarius~\cite{Springel:2008cc} and Via Lactea II~\cite{Diemand:2009bm} have each identified expansive populations of subhalos residing within the halos of Milky Way-like galaxies. Aquarius, for example, resolves their subhalo population down to masses of $3.24\times 10^4 M_{\odot}$~\cite{Springel:2008cc}. We can use the results of such simulations (extrapolated to include sub-resolution subhalos) to estimate how many dark matter subhalos should be detectable as gamma-ray sources by Fermi, as a function of the dark matter particle's mass and annihilation cross section.  

Following the approach described in Ref.~\cite{Berlin:2013dva}, updated to implement the mass-concentration relationship of Ref.~\cite{Sanchez-Conde:2013yxa}, we calculate the number of subhalos predicted to be detectable by Fermi. For a 34 GeV dark matter particle annihilating to $b\bar{b}$, for example, we predict the following number of high-latitude ($|b|>20^{\circ}$) subhalos that generate a gamma-ray flux ($>1$ GeV) greater than $F_{\rm threshold}$:
\begin{equation}
N \sim 1.2 \times \bigg(\frac{\sigma v}{10^{-26} \, {\rm cm}^3 \, {\rm s}^{-1}}\bigg)^{1.5} \, \bigg(\frac{F_{\rm threshold}}{10^{-9} \, {\rm cm}^{-2} \, {\rm s}^{-1}}\bigg)^{-1.5}.
\end{equation}
%
%
Although there exist non-negligible uncertainties regarding the distribution of subhalo concentrations, and the degree to which their outer mass is tidally stripped, reasonable variations in these parameters change the predicted number of detectable subhalos by only a factor of a few or less, and we consider our estimate to represent a reasonable prediction (the authors of Ref.~\cite{Schoonenberg:2016aml}, for example, arrive at a number of observable subhalos that is a factor of a few lower than our estimate). For an annihilation cross section near the upper limit derived from the observations of dwarf spheroidal galaxies~\cite{Drlica-Wagner:2015xua,Geringer-Sameth:2014qqa}, we expect Fermi to detect roughly one subhalo with $F_{\rm threshold} > 10^{-9} \, {\rm cm}^{-2} \, {\rm s}^{-1}$, and perhaps as many as $\sim$10 with $F_{\rm threshold} > 10^{-10} \, {\rm cm}^{-2} \, {\rm s}^{-1}$. If 3FGL J2212.5+0703 is in fact a dark matter subhalo (and none of the other 11 subhalos candidates are), it would suggest an annihilation cross section of $\sigma v \sim (0.12-2.5) \times 10^{-26}$ cm$^3$/s (90\% CL, statistical uncertainties only). Of course, other candidate sources could also be dark matter subhalos. In particular, several of the subhalo candidates listed in Table~\ref{candidatesummary} exhibit spectral shapes that are compatible with that observed from 3FGL J2212.5+0703 (and from the Galactic Center excess). If any of these sources are in fact subhalos, it would increase our estimate for the dark matter's annihilation cross section.

\begin{figure}
\begin{center}
\includegraphics[width=0.49\linewidth]{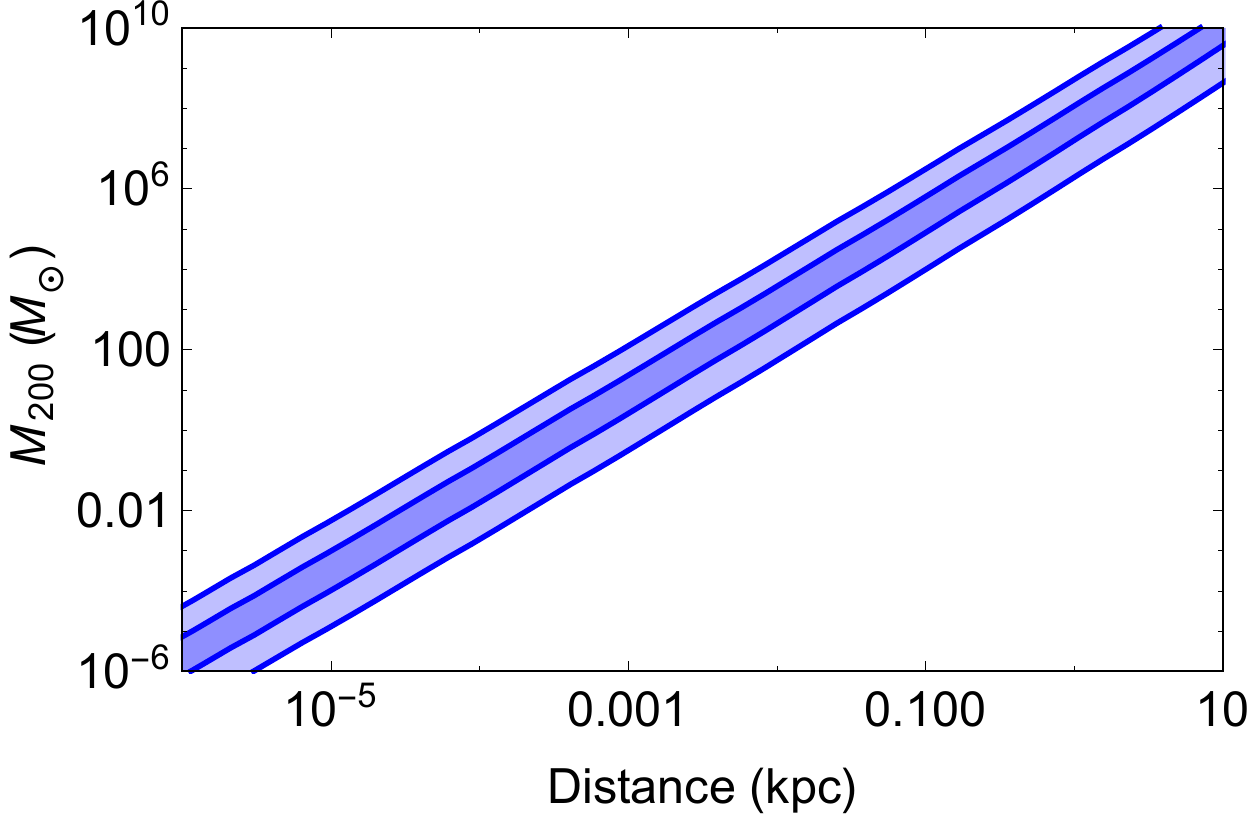}
\includegraphics[width=0.49\linewidth]{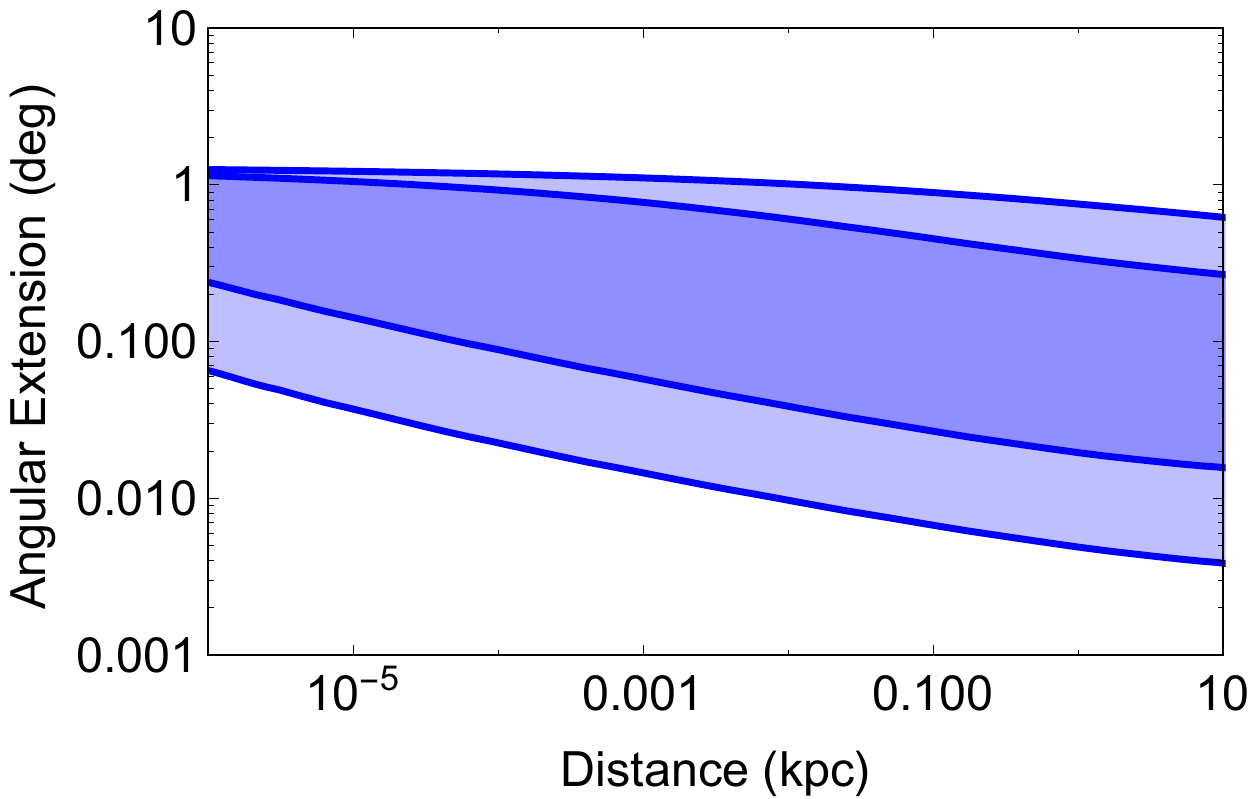}
\end{center}
\caption{Left frame: The mass (prior to tidal stripping) as a function of the distance to a subhalo that produces a gamma-ray flux equal to that observed from 3FGL J2212.5+0703. Right frame: The angular extension, $\sigma$, of the gamma-ray signal from a subhalo that produces a gamma-ray flux equal to that observed from 3FGL J2212.5+0703, as a function of the distance to the subhalo. In each frame, the blue bands reflect a range of values for the subhalo's concentration and mass loss due to tidal stripping (see text for details). Here, we have assumed a dark matter mass of 34 GeV and an annihilation cross section of $\sigma v =2\times10^{-26}$ cm$^{3}$/s to $b\bar{b}$.}
\label{distance}
\end{figure}


The gamma-ray flux and angular extent of 3FGL J2212.5+0703 can be used to constrain the mass and distance of the corresponding dark matter subhalo.  In the left frame of Fig.~\ref{distance}, we plot the mass of a subhalo (prior to tidal stripping) that produces the gamma-ray flux of 3FGL J2212.5+0703, as a function of distance. Here, we have assumed a dark matter mass of 34 GeV and an annihilation cross section of $\sigma v =2\times10^{-26}$ cm$^{3}$/s to $b\bar{b}$.\footnote{The left frame of Fig.~\ref{distance} can be adjusted to reflect any value of the cross section by shifting the distance scale by a factor of $[\sigma v/(2\times10^{-26} {\rm cm}^{3}{\rm s}^{-1})]^{1/2}$.} From the flux alone, one cannot disentangle the mass of a subhalo from its proximity. From the information contained in this plot, 3FGL J2212.5+0703 could equally well be a very large subhalo (perhaps even an ultra-faint dwarf galaxy) located at a distance of $\sim$$10$ kpc, or a solar mass clump of dark matter located within a parsec or so of the Solar System.\footnote{3FGL J2212.5+0703 is located within the region of the sky covered by the Sloan Digital Sky Survey (SDSS). If this object is a very massive subhalo hosting an ultra-faint dwarf galaxy, the stellar population might be identifiable from within this dataset. Although no evidence for a such a population has been found thus far~\cite{Balbinot:2012sn,Kim:2015xoa}, further investigation focusing on the direction of 3FGL J2212.5+0703 would be well motivated.}

Although one might imagine that the angular extension of a subhalo could be used to constrain its characteristics, there is an approximate degeneracy between size and distance that makes this measurement only mildly informative. In the right frame of Fig.~\ref{distance}, we plot the angular extension predicted for a subhalo with the same flux as 3FGL J2212.5+0703 as a function of distance, again assuming a dark matter mass of 34 GeV and an annihilation cross section of $\sigma v =2\times10^{-26}$ cm$^{3}$/s to $b\bar{b}$. In each frame of Fig.~\ref{distance}, the blue regions reflect the results for a range of concentration and tidal stripping assumptions. For both the inner dark blue and outer light blue bands, we have allowed the degree of tidal stripping to vary between 90\% and 99.5\%. The inner blue (outer blue) bands allow for variations in the subhalo's concentration by within a factor of 1.7 (2.9) around the central values presented in Ref.~\cite{Sanchez-Conde:2013yxa}. Although this plot illustrates a trend in favor of less extension for more distant (and more massive) subhalos, for the degree of extension observed from 3FGL J2212.5+0703 ($\sigma \approx 0.25^{\circ}$), it is not possible to meaningfully constrain its distance.

Even if the flux and angular extent of 3FGL J2212.5+0703 cannot definitely determine the mass of or distance to any corresponding dark matter subhalo, we can use the subhalo population found in numerical simulations to make probabilistic statements regarding its characteristics. In Ref.~\cite{Berlin:2013dva}, it was found that the differential number of subhalos detectable by Fermi is described by the following:
\begin{equation}
M \frac{dN}{dM} \propto M^{0.17},  \,\,\,\,\,\,\,\,   M<M_{\rm max},
\end{equation}
where $M$ is the subhalo mass.  From this expression, we find that half of all observable subhalos will have a mass within a factor of 60 of the maximum mass (where the maximum subhalo mass is that corresponding to the boundary between baryon-free subhalos and star-forming dwarf galaxies, on the order of $M_{\rm max} \sim 10^9 - 10^{10}\,M_{\odot}$, prior to tidal stripping). Furthermore, approximately 90\% of all detectable subhalos are predicted to be within a factor of $10^6$ of the maximum mass.  This exercise suggests that, if 3FGL J2212.5+0703 is a dark matter subhalo, then it is likely to have a mass of roughly $\sim$\,$10^3 - 10^{10} \, M_{\odot}$ (corresponding to $\sim$\,$10^1 - 10^8 \, M_{\odot}$ after tidal stripping), and reside at a distance between $\sim$\,$10$ pc and 10 kpc from the Solar System.

\section{Discussion and Summary}
\label{sec:conclusions}

In this paper, we have studied the spectrum and morphology of the emission from the unassociated gamma-ray source 3FGL J2212.5+0703, which we had previously identified as being apparently spatially extended.  Spatial extension is a feature expected of nearby dark matter subhalos, but not of most other classes of astrophysical gamma-ray sources. Moreover, as astrophysical sources that are capable of generating spatially extended gamma-ray emission are also invariably bright at other wavelengths, the detection of a gamma-ray source that was both unambiguously spatially extended and lacking in multi-wavelength counterparts would constitute a smoking gun for annihilating dark matter.

After studying the gamma-ray emission from a collection of 12 subhalo candidates, we have identified significant evidence of spatial extension for the source 3FGL J2212.5+0703. We find that the data prefers the morphology of a spherical (elliptical) dark matter subhalo with a tidally truncated Navarro-Frenk-White profile over that of a point source at the level of 4.7$\sigma$ (5.1$\sigma$).

An alternative explanation for the extension of 3FGL J2212.5+0703 is that it may be two or more gamma-ray sources located very nearby each other on the sky. In such a scenario, multiple sources could potentially be misidentified as a single extended source. Although we find that it is somewhat unlikely that a pair of bright, unassociated, high-latitude gamma-ray sources would lie this close to one another (corresponding to a probability of approximately two percent), this appears to be the least unlikely non-dark matter explanation for the observed morphology of 3FGL J2212.5+0703. In particular, we find that the inclusion of an additional gamma-ray point source in the fit, at the location of a known radio source, can accommodate the data approximately as well as a model with a single extended source. The current dataset is thus unable to distinguish between 3FGL J2212.5+0703 being a single extended source, or a pair of nearby point sources.

If 3FGL J2112.5+0703 is a dark matter subhalo, it would imply a dark matter mass of 18.4-32.7 GeV and an annihilation cross section of $\sigma v \sim 10^{-26}$ cm$^3/$s (for the representative case of annihilations to $b\bar{b}$).  Although the information available does not allow us to the determine the mass of or distance to this subhalo, simulations suggest that the first gamma-ray detected subhalos are likely to have masses in the range of $\sim$\,$10 \, M_{\odot}$ to $10^8\, M_{\odot}$. Thus 3FGL J2212.5+0703 could plausibly be a dark matter subhalo on the scale of an ultra-faint dwarf galaxy located at a distance of $\sim$$10$ kpc, or a much smaller clump of dark matter residing within a few tens of parsecs of the Solar System.

Within the 3FGL catalog, there are 12 bright ($F_{\gamma}>10^{-9}$ cm$^{-2}$ s$^{-1}$) sources located at high galactic latitudes ($|b|>20^{\circ}$) without known multi-wavelength associations. Most of these sources exhibit a spectral shape that is similar to that observed from 3FGL J2212.5+0703 (and the Galactic Center gamma-ray excess~\cite{Daylan:2014rsa,Calore:2014xka,TheFermi-LAT:2015kwa}). We consider this short list of sources to be particularly promising dark matter subhalo candidates, worthy of further study and observation.

Although the results of this paper are consistent with an interpretation of \\ 3FGL J2212.5+0703 as a dark matter subhalo, we emphasize that we cannot exclude the possibility that multiple nearby sources may instead be responsible for the apparent extension of this emission. With this ambiguity in mind, we implore the broader observational community to assist in clarifying the nature of this source (and of the other prospective dark matter subhalo candidates described in this paper), through the continued pursuit of deep and high-resolution multi-wavelength observations.

\bigskip
\bigskip

\textbf{Acknowledgments.} We would like to thank Alex Drlica-Wagner, Keith Bechtol, Frank Schninzel, Elizabeth Ferrara, Fernando Camilo, Paul Ray and Eric Charles for valuable discussions. BB is supported by the US Department of Energy Office of Science Graduate Student Research (SCGSR) Program under Contrast No. DE-AC05-06OR23100 and the US Department of Energy under Contract Nos. DE-FG02-00ER41132 and DE-SC0011637. DH is supported by the US Department of Energy under contract DE-FG02-13ER41958. Fermilab is operated by Fermi Research Alliance, LLC, under Contract No. DE-AC02-07CH11359 with the US Department of Energy. TL is supported by the National Aeronautics and Space Administration through Einstein Postdoctoral Fellowship Award No. PF3-140110. We acknowledge the University of Chicago Research Computing Center and the Ohio Supercomputer Center for providing support for this work.

\bibliography{J2212-Feb22.bib}
\bibliographystyle{JHEP}

\end{document}